\documentstyle[epsf]{article}
\begin{document}
\title{Locality, QED and Classical Electrodynamics}
\author{Dirk Kreimer\thanks{dirk.kreimer@uni-mainz.de}\\
Mainz University}
\maketitle
\begin{abstract}
We report on some conceptual changes in our present
understanding of Quantum Field Theory and muse about
possible consequences for the understanding of
$v>c$ signals.
\end{abstract}
\section{Introduction}
This paper is a musing about the role which locality might have to play in
the understanding of the phenomena which we gathered to discuss at this
conference. Locality has always been the crucial property in one's
approach to Quantum Field Theory (QFT). Mainly, there are
two lines of thought. One is to maintain locality. One then can follow
the standard derivation of Feynman rules and obtains a QFT
which suffers from UV-divergences. These divergences unavoidably reflect
the presence of local quantum fields, operator-valued distributions
which, when we clumsily try to multiply them at coinciding
space-time points, produce these UV-divergences.


A QFT as popular as Quantum Electrodynamics is the archetypical
example of such a theory. Its merits are indisputable though.
No experimental fact which is in conflict with its
predictions (and with its natural successor at higher energies,
the Standard Model of particle physics) has yet  been found,
and the theory is rightly praised for the accuracy with which it
describes nature.


Nevertheless, the price is high. The theory is formulated in
form of a highly dubious perturbative expansion: not only
do we have to ignore that this expansion in the coupling
constant is by no means convergent (even its Borel transform is in
doubt), but we also are confronted with the fact that
at each order in this expansion we have to manage ill-defined
mathematical expressions, due to the
very presence of UV-divergences, and thus
being a direct consequence of working with a local
QFT. Physicists found a way how to handle
these ill-defined expressions, known as renormalization.
Renormalization is often regarded  at best a technical
rather than an illuminating device. 


But their guilty conscience, for working with a priori
ill-defined quantities, often let physicists to abandon local
QFT, by either turning away completely from these issues, or,
and this is the other line of thought,
trying
to seek recourse in other solutions.
Most prominent here is string theory and its many
generalizations, which all replace the point-particle concept by somewhat
extended objects, and henceforth avoid UV divergences. Unfortunately,
these approaches so far fail dramatically to describe nature.


Thus, it is no surprise that a theory like QED was rarely 
taken as a guide in subtle conceptual questions, its own foundations seemingly being
built on unsecure ground.


This seems  to change  recently. 
What we see emerging
from the art of calculation of Feynman diagrams is the mathematics
which enables us to reconcile the practical successes of local QFT
with a mathematical foundation which is well-defined
and promising.


In the next section, I will describe some of the features of this
mathematical backbone of a local
QFT and in the final section draw some conclusions
concerning the topic of this workshop.


But before I stretch your patience with a rush through the
math of local QFT, let me first give you a hint how a QFT might have
something to say of relevance for what Prof.~Low called the Nimtz
anomaly 
\cite{Nimtz} in
his talk.


What Low showed us was how one can establish a simple
calculus which correctly describes the data delivered by the Nimtz
experiment
\cite{Low}.
His calculus showed that the Nimtz anomaly is completely determined
by height and length of the barrier, and showed the Nimtz anomaly as a function
of these two variables.


To have a manageable calculus, Low did a step which is everyday practice for someone
working in local Quantum Field Theory: he eliminated 
negligible contributions
which stem from unmeasurable internal high frequencies, 
a step which was mandatory
to arrive at the desired calculus, and 
which was justified by the fact that the
abandoned contributions were small with repect to the problem in question.
Nevertheless it is far reaching, 
as such tiny modifications  can have 
drastical conceptual consequences.


To me, it seems that it is indeed a well-posed question to ask
what the precise message is which local QFT has in store for
us in the context of conceptualization of notions like
signal, frequency band limitations and so on. Admittedly, 
one can regard classical field theory as detached from local QFT
for such classical  phenomena as the propagation of a microwavesignal to a 
(quenched) waveguide. 


But then, classical electrodynamics ought to be the limit of
QED in  some yet to be defined sense, and thus, if QED lives
in a universe in which a frequency band limitation, more usually dubbed
UV cut-off in the jargon of a practitioner of QFT,
is unavoidable, it might be that there is a message
which QFT has indeed in store for us: to be careful to find
those solutions to the classical equations of motions which remain stable under
negligible modifications of internal (unobserved) high frequency
components.

So what is then the conceptual backbone which underlies local QFT
and its need for a cut-off at high frequencies, which became
covariant renormalization in modern terminology?
\section{The concept of renormalization}
The basic idea is to trace back the concept of renormalization
to the structure of a Hopf algebra. In so doing, one recovers that
the process of renormalization has a well-defined mathematical
meaning and is related to the study of the diffeomorphism group
of spacetime \cite{hopf}.

\begin{figure}
\epsfysize=13cm
\epsfbox{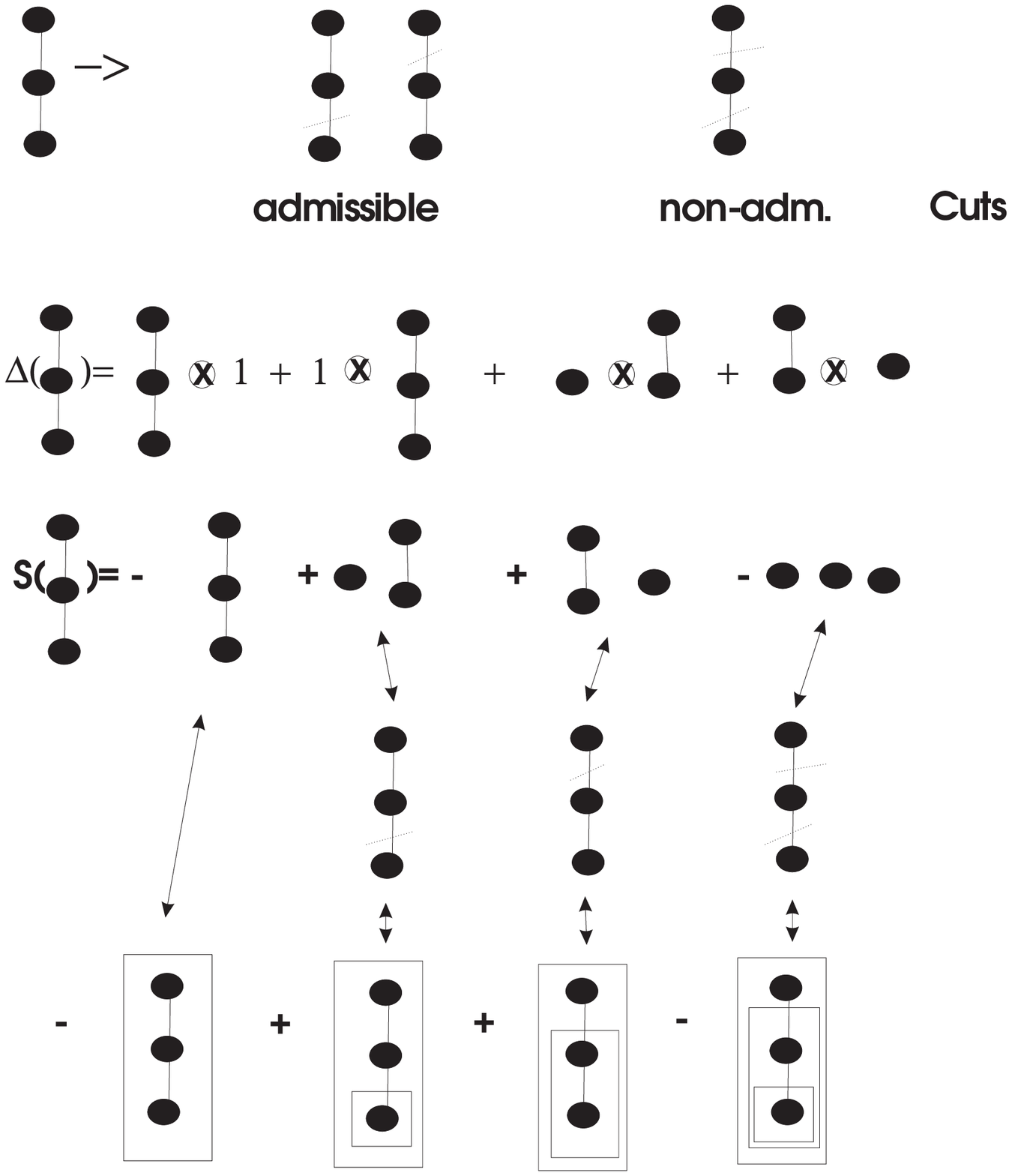}
\caption{The Hopf algebra of rooted trees. We define it using admissible
cuts on the trees, and give the coproduct $\Delta$  in terms of admissible
cuts. An admissible cut allows for at most one single cut
in any path from any vertex to the root \cite{hopf}.}
\label{f3}
\end{figure}

Fig.(\ref{f3}) summarizes some basic
notions. Especially, we indicate how to calculate the coproduct of a
rooted
tree. In so doing, we cut the tree in pieces. Each of these pieces
will
correspond to a local counterterm in an analytic expression
which renders a certain subdivergence finite.

In particular, this Hopf algebra has an antipode, which maps to
Feynman diagrams as their local counterterm. 

\begin{figure}
\epsfysize=16cm\epsfbox{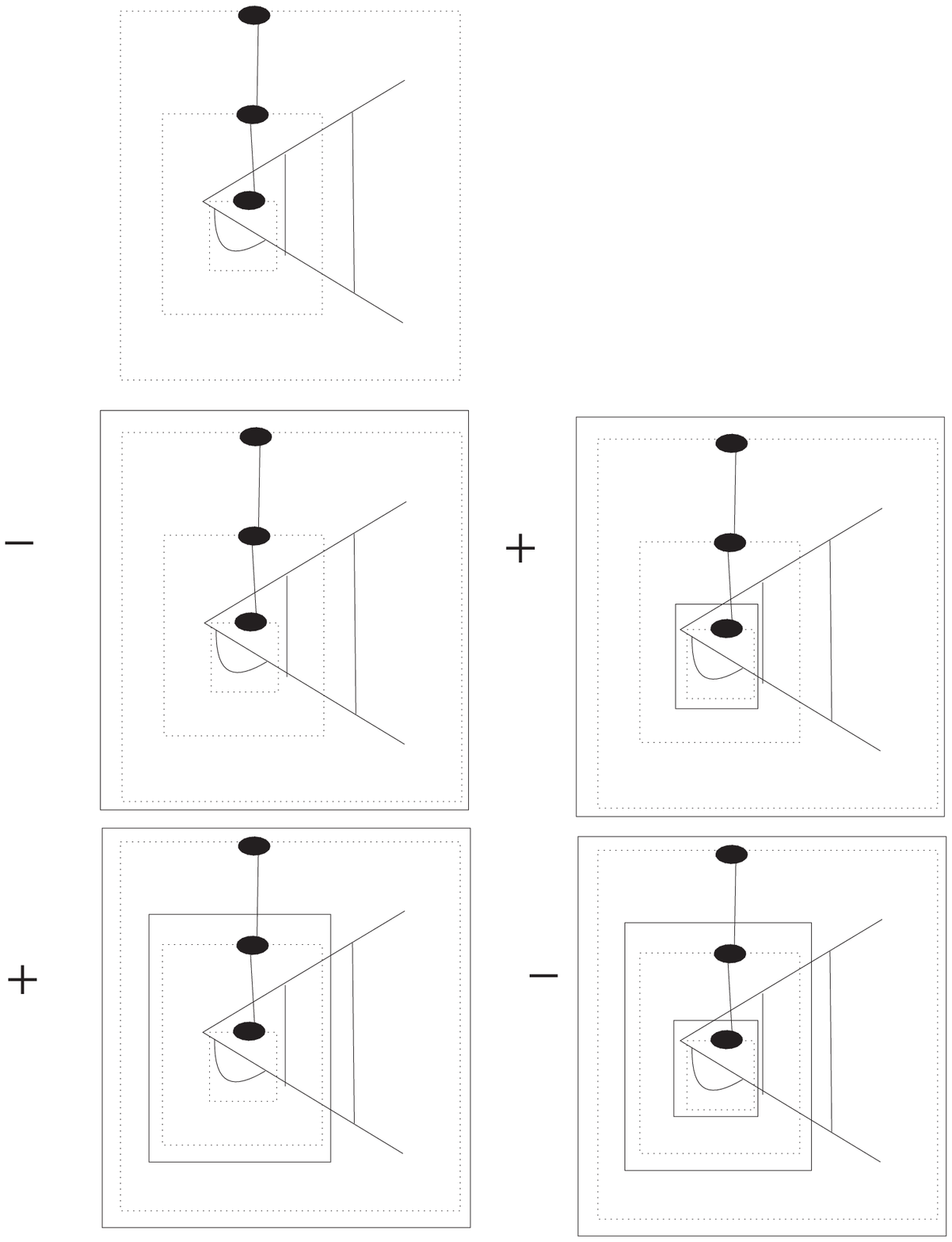}
\caption{The Hopf algebra of renormalization. We indicate how to
  assign a decorated tree
to a diagram. On such trees we establish the above Hopf algebra
structure.
Each black box corresponds to a cut on the tree, and these cuts
are in one to one correspondence with the forest structure.
We calculate the antipode on the tree, and represent the results
on Feynman diagrams, to find that the antipode corresponds to
the local counterterm.}\label{f4}
\end{figure}

Fig.(\ref{f4}) gives a diagrammatic explanation of this Hopf algebra, 
and shows how the combinatorics of the renormalization
procedure derives from 
this Hopf algebra.

The universal structure 
of the Hopf algebra \cite{hopf}  
guarantees that it can be applied as long as we can render a theory
finite by local counterterms: locality of the theory and its
Hopf algebra structure are two sides of the same coin.

Let us also stress that the appearance of this Hopf algebra structure
in QFT establishes a link to recent developments in mathematics.
Quite the same Hopf algebra turned up in the work of Alain Connes
and Henri Moscovici on the noncommutative
index theorem. The relation is by now made precise, and puts local
QFT on a firm mathematical ground \cite{hopf}.

The renormalized Green function can be recovered by the map
\[
\Gamma\to m[(S_R\otimes id)(\phi\otimes\phi)\Delta(T_\Gamma)].
\]
This map associates to the Feynman graph $\Gamma$ the renormalized
Feynman integral \cite{hopf}.


Indeed, this map can be written as
\begin{equation}
\Gamma\to\Gamma-\tau_R(\Gamma)+(id-\tau_R)\left[
\sum_{\gamma\subset\Gamma}Z_\gamma \Gamma/\gamma\right],
\end{equation}
where 
\begin{equation}
S_R(\phi(T_\gamma))\equiv
Z_\gamma=-\tau_R(\gamma)-\tau_R\left[\sum_{\gamma^\prime\subset\gamma}
Z_{\gamma^\prime}\gamma/\gamma^\prime\right],
\end{equation}
and this map is induced by the antipode
\begin{equation}
S[T_\gamma]=-T_\gamma-\sum_{\gamma^\prime\subset\gamma}S[\gamma^\prime]
\gamma/\gamma^\prime.
\end{equation}
Hence, in accordance with \cite{hopf} we find the $Z$-factor of a
graph $\gamma$ as derived from the antipode in the Hopf algebra
of rooted trees.

\section{Conclusions}
What is now the message which we can learn from this structure
of perturbative QFT? It seems to me that the only message is 
that we shall be not
so certain in which function space to look for classical solutions.
The above described mathematics relate local QFT to functional
analysis and noncommutative geometry.
The question to which extent undetected high frequency components 
are allowed to contribute is
of direct relevance for the space of solutions considered, and a mathematical
rigorous treatment will have to explore the mathematics
of functional analysis and operator algebras, the very disciplines in which
a local QFT has its basis anyhow. 
The correct asymptotic behaviour of a local QFT, 
characterized by the unimportance of high frequency
components, will have meaning  for the limit
to classical field theory.
A quantification of this statement and hence a derivation
of the Nimtz anomaly will have to wait
until our understanding of QFT has reached maturity. 
\section*{Acknowledgements}
It is a pleasure to thank the organizers of the 
$v>^?c$
workshop for support and hospitality.

\end{document}